\documentclass[aps,twocolumn,showpacs,floatfix]{revtex4}

\usepackage{amsmath}

\usepackage{graphicx}

\begin{document}

\title{Tunneling and Magnetic Characteristics of Superconducting ZrB$_{12}$ Single Crystals}
\author{M.I. Tsindlekht, G.I. Leviev, I. Asulin, A. Sharoni, O. Millo and I. Felner}
\affiliation{The Racah Institute of Physics, The Hebrew University
of Jerusalem, 91904 Jerusalem, Israel}
\author{Yu.B. Paderno and V.B. Filippov }
\affiliation{Institute for Problems of Materials Science, National
Academy of Sciences of Ukraine, 03680 Kiev, Ukraine}
\author{M.A. Belogolovskii}
\affiliation{Donetsk Physical and Technical Institute, National Academy of Sciences of
Ukraine, 83114 Donetsk, Ukraine}

\date{\today}

\begin{abstract}
Bulk and surface properties of high-quality single crystals of
zirconium dodecaboride have been studied in the temperature range
from 4.5 K up to the superconducting transition temperature which
is found to be nearly 6.06 K. Scanning tunnelling spectroscopy
data, together with dc and ac magnetization measurements, are
consistent with the conventional $s$-wave pairing scenario,
whereas they disagree in estimates of the electron-phonon coupling
strength. We explain the divergence, supposing a great difference
between the surface and bulk superconducting characteristics of
the compound. This assertion is supported by our findings of a
non-linear magnetic response to an amplitude-modulated alternating
magnetic field, testifying to the presence of surface
superconductivity in the ZrB$_{12}$ samples at dc fields exceeding
the thermodynamic critical field.
\end{abstract}
\maketitle


Due to the unique combination of physical properties such as: high
melting point, hardness, thermal and chemical stability
metal-boron compounds have found different applications
~\cite{BOR,SER}. The discovery of superconductivity in metallic
MgB$_2$ at the unexpected $T_{\rm{c}}$ of 39 K~\cite{AKIM} has
caused a great interest in other transition-metal diborides which
show lower $T_{\rm{c}}$ values. Related experimental efforts have
recently been  extended to a wider class of binary
boron-containing intermetallic compounds, in particular, to
zirconium dodecaboride~\cite{PAD2,DAG}.

First measurements on polycrystalline ZrB$_{12}$ (as well as on
many other boron-rich compounds) were done by Matthias \textit{et
al.} in~\cite{MAT} where $T_{\rm{c}} =5.82$ K was determined from
the sharp transition in specific heat data. (Our estimate for
$T_{\rm{c}}$ detected by dc magnetization measurements agrees with
the value of $T_{\rm{c}}\approx 6.0$ K~\cite{HAM}). Despite
assertions that the ZrB$_{12}$ phase exists only at high
temperatures~\cite{POST}, large high-quality single crystals of
ZrB$_{12}$ were grown in Kiev~\cite{PAD} and by Leithe-Jasper
\textit{et al.}~\cite{LEIT}. Whereas this work~\cite{LEIT}
reported only crystallographic data for the crystals, two
groups~\cite{PAD2,DAG} have performed investigations of ZrB$_{12}$
single crystals grown in Kiev~\cite{PAD} by different physical
methods. Some unexpected results have been derived and different
conclusions relating to the nature of superconductivity in this
compound were claimed. Gasparov et. al.~\cite{PAD2} observed a
linear temperature dependence of the magnetic field penetration
depth below 3 K, which contradicts the standard BCS theory, and
proposed a $d$-wave-like pairing in ZrB$_{12}$. On the other hand,
Daghero \textit{et al.}~\cite{DAG}, based on resistivity vs.
temperature data and the well-known McMillan formula for
$T_{\rm{c}}$, concluded that ZrB$_{12}$ is a conventional $s$-wave
superconductor with a zero-temperature energy-gap value $\Delta
(0)=0.97$ meV and the ratio $2\Delta (0)/T_{\rm{c}}=3.64$. The
same authors~\cite{DAG} have also measured point-contact
conductance spectra that were found to be dominated by features
typical of an $s$-wave superconductor but with the value $\Delta
(T)=0.97$ meV at $T$ nearly 4 K (the BCS model yields
$2\Delta(0)/T_{\rm{c}}=4.8$). The only issue that the two
groups~\cite{PAD2,DAG} agree on, is that the single crystals
studied are type-II superconductors with an upper critical field
$H_{\rm{c2}}(0)$ above 1000 Oe, as estimated in
Ref.~\onlinecite{PAD2} from resistance measurements, and in
Ref.~\onlinecite{DAG} as the field in which the Andreev-reflection features
in conductance characteristics disappear.

In this work, we have studied tunneling and magnetic
characteristics of ZrB$_{12}$ single crystals grown in Kiev. The
measurements were performed for temperatures ranging from above
$T_{\rm{c}}$ down to 4.2 K, where analytical expressions obtained
within the standard Ginzburg-Landau approximation~\cite{AA} may be
applied. In the following, we address two issues: the pairing
symmetry in ZrB$_{12}$ single crystals and whether they are really
type-II superconductors. We have also studied surface
characteristics with linear and non-linear ac magnetic response
measurements. Chemically, dodecaborides are the extremely stable
materials in comparison with other borides and are characterized
by a strong surface resistance to mechanical and chemical
factors~\cite{YUK}. The difference between surface and bulk
properties is discussed.

Large rods of ZrB$_{12}$ single crystals were grown by the Kiev
group with typical dimensions of about 6 mm in diameter and up to
40 mm in length. A $10.3\times 3.2\times 1.2$ mm$^3$ rectangular
sample was cut from the rod and was polished mechanically by
diamond and chemically etched in a boiling HNO$_3$/H$_2$O $1:1$
mixture for 10 min to remove the Beibly layer. Room temperature
X-ray diffraction measurements performed in Kiev and in Jerusalem
confirmed that the ZrB$_{12}$ sample is a single-phase material
with a UB$_{12}$ structure (the space group of $Fm3m$, $a=7.407$
\AA~\cite{KEN}).

The order parameter symmetry of ZrB$_{12}$ was studied by scanning
tunneling spectroscopy measurements of current $I$ - voltage
$V$ curves. In contrast to the point-contact technique~\cite{DAG},
the differential conductance $dI/dV$-versus-$V$ yields direct
information on the local quasiparticle density of states, and
hence, on the superconductor gap symmetry. The samples were
carefully cleaned with ethanol in an ultrasonic bath just before
they were mounted into our cryogenic home-made scanning tunneling
microscope~\cite{OD}. Before inserting He exchange gas (through a
trap) the sample space was evacuated and the device was dipped
into a liquid-helium storage dewar. After sufficiently long
thermalization period the sample temperature was somewhat above
4.2 K, but lower than 4.4 K, as was indicated by a sensor placed
nearby. Tunneling measurements were performed for junction
normal-state resistances between 50 and 500 MOhm. $I$-$V$
characteristics were differentiated numerically in order to obtain
normalized spectra $(dI/dV)_S/(dI/dV)_N$, the ratios of
differential conductances in
superconducting and normal states. The
spectra were compared with a temperature-smeared~\cite{BGS}
version of the Dynes formula~\cite{DYN} that takes into account
the effect of incoherent scattering events inside a
superconductor, by introducing a damping parameter $\Gamma $ into
the conventional $s$-wave BCS expression for a normalized
quasiparticle density of states $N(\varepsilon
)=\mathrm{Re}[(\varepsilon -i\Gamma )/ \sqrt {(\varepsilon
-i\Gamma )^2-\Delta (T)^2}]$.

\begin{figure}[tbp]
\includegraphics[width=0.48\textwidth]{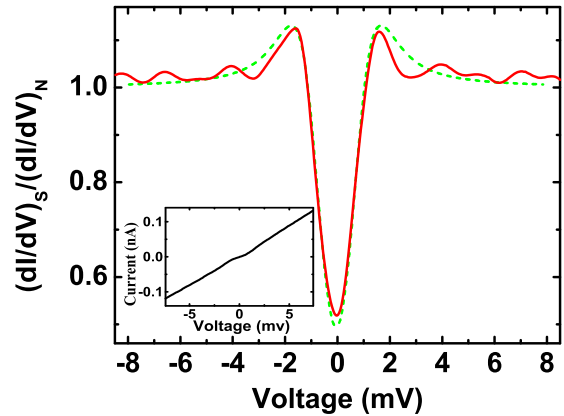}
\caption{(Color online) Representative tunneling spectrum of
ZrB$_{12}$ at $T$ nearly 4.2 K (solid line) together with its fit
to the temperature-smeared Dynes formula (see the text) shown by a
dashed line, with fitting parameters $\Delta (T)$=0.97 meV and
$\Gamma $=0.15 meV. The inset shows the initial current-voltage
characteristic.} \label{f1}
\end{figure}

The well reproducible local tunneling characteristics
(Fig.~\ref{f1}) did not change significantly with the tip position
and/or the device settings. All curves demonstrated coherence
peaks with a pronounced minimum at $V=0$ and a near-gap structure
symmetrical with respect to the bias voltage $V$. Gap values
$\Delta(T\approx4.2 K)$ were found to be 0.97$\pm$0.01 meV (which
coincides well  with point-contact findings~\cite{DAG} for the
same temperature) with $\Gamma$ not exceeding 0.15 meV. A
relatively small value of the damping factor (in particular,
compared with those in Ref.~\onlinecite{DAG}) is believed to prove
the high quality of the sample surface. For all tip positions we
have not observed any signs of a zero-bias peak known as a
fingerprint of the $d$-wave pairing~\cite{TAN} and, hence, may
reject the assumption about an unusual symmetry of the order
parameter in ZrB$_{12}$.

\begin{figure}[tbp]
\includegraphics[width=0.48\textwidth]{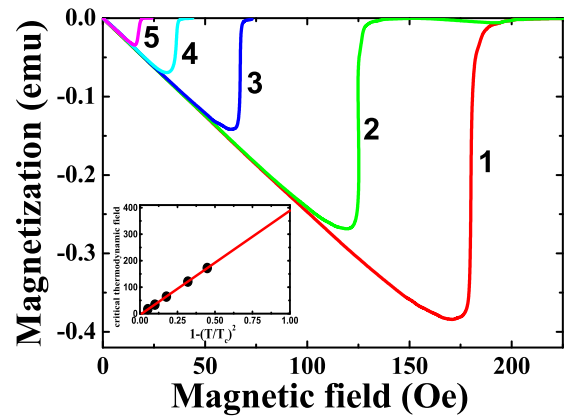}
\caption{(Color online) Magnetic field dependence of the magnetic
moment at various temperatures: 1 - 4.5 K; 2 - 5.0 K; 3 - 5.5 K; 4
- 5.75 K; 5 - 5.9 K. The solid line in the inset is an
extrapolation of the thermodynamic critical field behavior (2)
deduced from the calculated $H_{\rm{c}}(T)$ shown by circles.}
\label{f2}
\end{figure}

The latter statement was also confirmed by zero-field-cooling dc
magnetization $M$-versus-$H$ measurements performed using a
commercial SQUID magnetometer (Quantum Design MPMS) with external
magnetic fields aligned parallel to the long axis of the sample.
In the normal state (above 6 K) the dc susceptibility
$\chi_{\rm{dc}}=M/H$ is diamagnetic as expected ~\cite{HAM}.
Fig.~\ref{f2} shows the magnetization curves measured at various
temperatures, corrected by the demagnetization factor $N$.
Usually, when the external field is parallel to the long axis
of a parallelepiped, $N$ is small (the reduction in the $H$ is a
few percent) and can be neglected. However, for a sharp transition
(as in our case) its width is of the same order as this reduction
and more detailed analysis is needed. Using the expressions for an
ellipsoid~\cite{BRA} and interpolating between the extreme cases
that correspond to two different transverse sizes, $N=0.06$ was
chosen. After this correction, the nearly vertical magnetization
curves were obtained as shown in Fig.~\ref{f2}.
This vertical shape close to the critical field values and the
tiny hysteresis loops observed (not shown in Fig.~\ref{f2}) lead
to the conclusion that bulk ZrB$_{12}$ is a type-I superconductor
or, alternatively, a type-II superconductor in which the
Ginzburg-Landau parameter $\kappa $ is slightly above the marginal
value of $\kappa $=0.71~\cite{AA}.

>From the data presented in Fig.~\ref{f2} we can estimate the
thermodynamic critical fields $H_{\rm{c}}$, whose equivalent
magnetic energy is equal to the area under the measured
$M$-versus-$H$ dependence
$\int_{0}^{\infty}[-M(H)]dH=H_{\rm{c}}^{2}/(8\pi )$. According to
the standard BCS theory for an $s$-wave superconductor, the
critical field curve $H_{\rm{c}}(T)$
should saturate at low temperatures and be linear in the vicinity
of $T=T_{\rm{c}}$~\cite{AA}:
\begin{equation}
\label{eq1}
H_{\rm{c}}(T)=1.735H_{\rm{c}}(0)(1-T/T_{\rm{c}}).
\end{equation}
The latter analytical result allows us to prove the conventional
symmetry of the pairing in ZrB$_{12}$, and to determine its
$T_{\rm{c}}$ value. Experimental data shown in Fig.~\ref{f2}
agreed very well with the linear dependence (\ref{eq1}) (the
derivative $(dH_{\rm{c}}(T)/dT)_{T=T_{\rm{c}}}$ was equal to -110
Oe/K) and an extrapolation of $H_{\rm{c}}(T)$ to zero yielded
$T_{\rm{c}}$ = 6.06 K. Now it is possible to use the empirical
formula
\begin{equation}
\label{eq2}
H_{\rm{c}}(T)=H_{\rm{c}}(0)[1-(T/T_{\rm{c}})^2],
\end{equation}
which interpolates the overall behavior of $H_{\rm{c}}(T)$ between
$T=0$ and $T=T_{\rm{c}}$, to estimate the zero-temperature
thermodynamic critical field. Fig~\ref{f2} (inset) yields
$H_{\rm{c}}(0)$ as 390 Oe. Using the density of states at the
Fermi energy calculated recently for ZrB$_{12}$ by Shein and
Ivanovskii~\cite{SHE} $N(\varepsilon _F)=1.687$ 1/(eV$\cdot $cell)
and the bulk energy-gap value of 0.97 meV from
Ref.~\onlinecite{DAG}, one can estimate the condensation energy
within the BCS theory and to evaluate the thermodynamic critical
field $H_{\rm{c}}(0)$. The value obtained is nearly 300 Oe which
compares reasonably well with our estimation of $H_c(0)$ listed
above. Again, we see no reason to suppose any deviation from the
conventional gap symmetry and the phonon origin of the Cooper
pairing in ZrB$_{12}$.

More problematic is the question about the electron-phonon
coupling strength. As was pointed out by Rammer~\cite{RAM}, the
ratio $H_{\rm{c}}(0)/(T_{\rm{c}}|dH_{\rm{c}}/dT|_{T=T_{\rm{c}}})$
can serve as its indicator. In the weak-coupling limit it equals
0.58, as follows from Eq.~(\ref{eq1}), whereas strong-coupling
effects lead to a significant reduction of this
quantity~\cite{RAM}. In our case, this ratio equals to 0.59 which,
together with the conclusions of the resistivity
measurements~\cite{DAG} places ZrB$_{12}$ into the category of
weak-coupling superconductors. At the same time, using the
temperature dependence of the BCS energy gap~\cite{AA} and our
value for the energy gap at $T\approx 4.2$ K, we obtain
$\Delta(0)=(1.21\div1.24)$ meV in excellent agreement with
$\Delta(0)=1.22$ meV extrapolated in Ref.~\onlinecite{DAG} from
point-contact measurements. In this case the ratio
$2\Delta(0)/T_{\rm{c}}$, the characteristic of the electron-phonon
coupling strength~\cite{CAR}, equals $4.75\pm0.10$, indicating
clearly an extremely strong electron-phonon interaction. According
to Ref.~\onlinecite{CAR}, there is a general trend of the
$2\Delta(0)/T_{\rm{c}}$ growth with increasing $T_{\rm{c}}/\omega
_{\rm{ln}}$ ($\omega _{\rm{ln}}$ is a characteristic energy of
lattice vibrations) and for $2\Delta(0)/T_{\rm{c}}$=4.75, $\omega
_{\rm{ln}}$ should be as great as 0.15~\cite{CAR}. The presence of
such low phonon energies is doubtful, and, to explain this result,
we assume that the order parameter is higher near the sample
surface. In particular, it should lead~\cite{FJ} to a surface
nucleation fields $H_{\rm{c}3}$ larger than those expected for a
uniform sample.

To study the superconducting sheath state in the field range
$H_{\rm{c}}<H<H_{\rm{c}3}$, we applied an additional small ac
magnetic field $h(t)$ upon the coaxial dc field $H$ and detected
the linear and nonlinear responses~\cite{ROLL}. The measurements
have been done with our original home-made setup~\cite{TM} adapted
to the MPMS SQUID magnetometer. In linear experiments the ac field
$h(t)=h_0\cos \Omega t$ with $0<h_{0}<0.4$ Oe and $\Omega /2\pi
=1455$ Hz was generated by a copper solenoid inside the
magnetometer and the ac susceptibility versus $H$ was measured by
the two-coil method. To study nonlinear characteristics of the
sample, the ac perturbation had a form of an amplitude-modulated
ac field $h(t)=h_{0}(1+\alpha \cos \Omega t)\cos\omega t$ with two
additional parameters $\alpha \approx 0.9$ and $\omega /2\pi =3.2$
MHz (see for details Ref.~\onlinecite{TM} where the same technique was
applied to a Nb single crystal) and the amplitude of a rectified
signal at the modulation frequency $A_{\Omega}$ was measured as a
function of the dc field $H$.

\begin{figure}[tbp]
\includegraphics[width=0.48\textwidth]{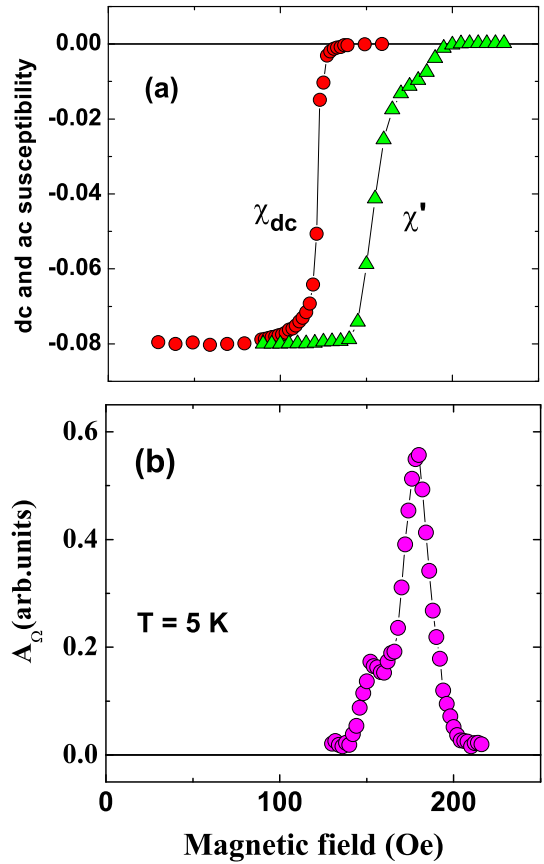}
\caption{(Color online) Magnetic field dependence of the dc
susceptibility $\chi_{\rm{dc}}$ compared with the real part of the
ac susceptibility $\chi^{\prime}$ at the fundamental frequency
$\Omega $ (a) and of the rectified signal amplitude $A_{\Omega }$
(b); $T=5.0$ K.} \label{f3}
\end{figure}
Fig.~\ref{f3}a exhibits a gradual shift of the real part
$\chi^{\prime}$ of the ac susceptibility compared with
$\chi_{\rm{dc}}$ calculated from dc magnetization curves. Such
behavior was explained
in Ref.~\onlinecite{ROLL} as an impact of the surface
superconductivity that appears in perfect samples in a dc field
parallel to the sample surface and persists in a surface region
adjacent to a vacuum interface up to a field $H_{\rm{c}3}$. Near
$T_{\rm{c}}$ $H_{c3}$ is defined by a simple relation
$H_{\rm{c}3}=2.38\kappa H_{\rm{c}}$~\cite{AA}. The shift between
ac and dc susceptibility curves strongly depends on the amplitude
and frequency of the ac field $h(t)$ and, because of that, any
interpretation and determination of a critical field value
$H_{\rm{c}}$ based on ac techniques should be done carefully. A
nonlinear response $A_{\Omega}$ was detected only above
$H_{\rm{c}}$ (Fig.~\ref{f3}b) where $\chi_{\rm{dc}}$ vanishes, and
hence, the sample bulk was in a normal state. It corresponds to
the direct observations~\cite{ROLL} of a nonlinear nature of the
response wavefront when a sinusoidal ac field is applied to a
specimen in a superconducting sheath state. Variations of $h_0$
causes an identical behavior of the $A_{\Omega}$ in a Nb single
crystal~\cite{TM} where the presence of surface superconductivity
was proven by different authors (see, for example,~\cite{KOT}). In
our case, the ratio $H_{\rm{c}3}/H_{\rm{c}}$ near $T_{\rm{c}}$ is
about 1.8. For a vacuum interface, the Ginzburg-Landau parameter
should exceed the marginal value of $\kappa $=0.71, which divides
type I and type II superconductors. But if the order parameter
increases near the interface (as it is argued above), then the
ratio of the surface nucleation field $H_{\rm{c}3}$ to the
thermodynamic critical field can be dramatically enhanced (see
Fig.~1 in Ref.~\onlinecite{FJ}). Then our
data could be interpreted with an
assumption of type-I superconductivity. Additional experiments are
needed to define the value of the Ginzburg-Landau parameter in
ZrB$_{12}$ single crystals.

In conclusion, we have proven experimentally that ZrB$_{12}$
single crystals are conventional $s$-wave superconductors with
enhanced surface characteristics. The latter statement can
principally explain the difference in estimations of the
electron-phonon coupling strength and of critical magnetic fields
between the three studies performed on the same single crystals.
More investigations are needed to explain all findings in this
non-trivial material, in particular, the absence of the field
direction effect reported
in Ref.~\onlinecite{DAG} (note that surface
superconductivity should be strongly suppressed when the magnetic
field has a component normal to the surface). We believe that
zirconium dodecarboride is an interesting and fruitful compound for
future experiments because of three reasons. First, due to
excellent surface properties it can serve as a very suitable model
material for studying specific near-surface superconducting
properties that are not yet well understood ~\cite{KOT}. Second,
our experiments show that it is an unusual marginal superconductor
near the border between type-I and type-II superconductors. And
last, ZrB$_{12}$ (similar to other dodecaborides) rises above
conventional materials due to its outstanding resistance to
external mechanical and chemical factors. We believe that its
unique material properties and comparatively simple
superconducting characteristics will attract the attention of the
applied physics community to the compound which can find its place
among various superconducting bulk applications where strong
abrasion- and chemical resistant properties are required.

We are thankful to R.S.~Gonnelli, I.R.~Shein and J.~K\"{o}tzler
for valuable discussions. The work was supported by the INTAS program
under the project No. 2001-0617.

\end{document}